\documentclass[useAMS,usenatbib]{mn2e}

\usepackage{ulem} 
\usepackage{graphicx}
\usepackage{amsmath}
\usepackage{ulem}
\usepackage{rotating}
\usepackage{amssymb}
\usepackage[pdfborder={0 0 0}]{hyperref}
\usepackage{aas_macros}

\newcommand{\hst}{{\sl HST}}

\def\lapprox{\hbox{\lower .8ex\hbox{$\,\buildrel < \over\sim\,$}}}
\def\gapprox{\hbox{\lower .8ex\hbox{$\,\buildrel > \over\sim\,$}}}

\title[Extending \textit{Gaia}\,DR2 with \textit{HST}\,astrometry. II.]{
Extending \textit{Gaia}\,DR2 with \textit{HST}\,narrow-field\,astrometry.\,II. Refining the method on WISE J163940.83$-$684738.6\thanks{
Based on observations with the NASA/ESA {\it Hubble
Space Telescope}, obtained at the Space Telescope Science Institute,
which is operated by AURA, Inc., under NASA contract NAS 5-26555.
}
}
\author[L.\,R.\,Bedin \& C.\ Fontanive]{
  L.\,R.\,Bedin$^{1}$\thanks{E-mail: luigi.bedin@oapd.inaf.it} and 
  C. Fontanive$^{2}$
\\  
$^{1}$INAF-Osservatorio Astronomico di Padova, Vicolo dell'Osservatorio 5, I-35122 Padova, Italy\\
$^{2}$Center for Space and Habitability, University of Bern, Gesellschaftsstrasse 6, 3012 Bern, Switzerland
}

\begin{document} 

\date{Accepted 2020 February 21. Received 2020 February 21; in original form 
2020 January 28}

\pagerange{\pageref{firstpage}--\pageref{lastpage}} \pubyear{202X}

\maketitle
 
\label{firstpage}

\begin{abstract}
%
In the second paper of this series we perfected our method of linking
high precision \textit{Hubble Space Telescope} astrometry to the
high-accuracy Gaia\,DR2 absolute reference system to overcome the
limitations of relative astrometry with narrow-field cameras.
Our test case here is the Y brown dwarf WISE\,J163940.83$-$684738.6,
observed at different epochs spread over a 6-yr time baseline with the
\textit{Infra-Red} channel of the \textit{Wide Field Camera 3}.
We derived significantly improved astrometric parameters compared to
previous determinations, finding:
$(\mu_{\alpha\cos{\delta}},\mu_\delta,\varpi)=$
($  577.21\pm0.24$\,mas\,yr$^{-1}$,
 $-3108.39\pm0.27$\,mas\,yr$^{-1}$,
$210.4\pm1.8$\,mas$)$. 
In particular, our derived absolute parallax ($\varpi$) corresponds to
a distance of 4.75$\pm$0.05\,pc for the faint ultracool dwarf.
\end{abstract}


%
\section{Introduction}
\label{introduction}
%
%

Distance is a crucial parameter for investigating the basic physical
properties of any astronomical object. Indeed, precise distances are
essential to connect measured properties to intrinsic characteristics
(e.g. apparent to absolute magnitude), and therefore to compare
observations to theoretical predictions.

Current atmospheric and evolutionary models struggle to reproduce the
photometric properties of the lowest-mass and coolest brown dwarfs
\citep{Schneider2016,Leggett2017}. Measurements of accurate distances
allow for the determination of absolute fluxes and unbiased spectral
energy distributions, making such measurements a necessary step to
improve characterisation and modelling of low-mass objects
(e.g. \citealp{Kirkpatrick2019}). Precise distance estimates can also
be used to compare the appearance of individual objects to
well-calibrated colour-magnitude diagrams. In particular, the
identification of outliers along the standardised locus can probe
secondary attributes of these substellar objects. For example,
over-luminous sources may be indicative of unresolved binarity
\citep{Manjavacas2013,Tinney2014,Kirkpatrick2019}. Likewise,
excessively red or blue colours can trace a deviant surface gravity or
metallicity, or be evidence for diverse atmospheric features like
clouds \citep{Knapp2004,Chiu2006,Cruz2007,Cruz2009}.

Finally, the study of well-defined and complete samples in space
allows for the development and testing of formation and evolution
theories (e.g. \citealp{Kirkpatrick2019}). Current observations of
substellar mass functions and space densities are in tension with
model predictions \citep{Burgasser2004,Allen2005,Pinfield2006,Kirkpatrick2012}.
High-confidence volume-limited samples can only be achieved through
measurements of distances, which are thus required to obtain a
comprehensive portrait of the local substellar population.

Parallaxes are the most direct measures of distance for stellar and
substellar objects. With the extensive sky coverage of large
astrometric missions (e.g. \textit{Gaia}, \textit{Hipparcos}), most
stars in the solar neighbourhood and nearby moving groups or
star-forming regions have reliable parallax measurements. Isolated
brown dwarfs and free-floating planetary-mass objects, on the other
hand, are generally too faint and too red to be detected by these
broad surveys, and very few substellar objects are typically included
in these astrometric catalogues.

Spectrophotometric distances (based on expected relations between
absolute magnitude and spectral type or apparent photometry) are often
the only viable way to estimate distances for intrinsically faint
objects. However, significant disagreements have been found between
model-derived spectrophotometric distances and parallactic
measurements (e.g. \citealp{Kirkpatrick2011,Kirkpatrick2012}), and the
former estimates are often viewed as unreliable
\citep{Cushing2011,Liu2011}.  Some dedicated programs aim at deriving
trigonometric parallaxes for brown dwarfs, such as the Hawaii Infrared
Parallax Program \citep{Dupuy2012,Liu2016} or the Brown Dwarf
Kinematics Project \citep{Faherty2012} (see also
\citealp{Dupuy2013,Manjavacas2013,Manjavacas2019,Martin2018,Kirkpatrick2019}
for other compilations of parallactic distances). Despite these
remarkable efforts, the typical precision reached in these
observationally-expensive campaigns results in substantial
uncertainties in the underlying distances, and large inconsistencies
remain between programs for the faintest targets
(e.g. \citealp{Beichman2014}).

We recently devised in \citet{Bedin&Fontanive2018} (hereafter,
Paper\,I) a new method to improve the astrometric precision of
\textit{Hubble Space Telescope (HST)} observations and derive
astrometric parameters with \textit{Gaia}-level precisions for sources
too faint to be detected with \textit{Gaia}. This provides a powerful
procedure to infer highly-precise distances for faint, ultracool brown
dwarfs. For our test case target, the Y1 brown dwarf
WISE~J154151.65$-$225024.9 \citep{Cushing2011,Schneider2015}, we
achieved a precision at the milli-arcsecond (mas) level on parallax
and at the sub-mas level on proper motion, improving by an order of
magnitude the uncertainties from previous estimates.

In this paper, we further improve our method and apply it to the Y
dwarf WISE~J163940.83-684738.6 in order to constrain its astrometric
parameters to unprecedented levels.

%
\section{W1639$-$6847} 
\label{target}
%

WISE~J163940.83$-$684738.6 (hereafter W1639$-$6847) was first reported
by \citet{Tinney2012}, after using ground-based methane imaging to
carefully resolve the near-infrared counterpart of a blended WISE
source. The authors estimated a Y0$-$Y0.5 spectral type based on
near-infrared spectroscopy. \citet{Tinney2014} subsequently found
W1639$-$6847 to show under-luminous $J$ and $W2$ absolute magnitudes
and to be more consistent with a later type of Y0.5. The authors also
concluded that some photometric properties of the brown dwarf were in
better agreement with Y1 brown dwarfs.  Using \textit{HST}
spectroscopy, \citet{Schneider2015} found that the $J$-band peak of
W1639$-$6847 matched well with the Y0 spectral standard, in agreement
with previous spectral type estimates. However, the $Y$-band peak
appeared to be significantly blue-shifted when compared to the T9
spectral standard, and $Y-J$ colour seemed unusual relative to other
Y0 dwarfs. This led \citet{Schneider2015} to classify W1639$-$6847 as
Y0-Peculiar (Y0pec), which is since the adopted spectral type of this
object.

\citet{Opitz2016} studied W1639$-$6847 as part of a multiplicity
survey, attempting to resolve close Y dwarf binaries with the Gemini
Multi-Conjugate Adaptive Optics System. The authors were able to rule
out secondary companions down to 3.5 mag fainter from separations
beyond 0.5 AU. However, the search for companions was limited to the
inner 2.5 AU around the primary. No search for wide binary companion
around W1639$-$6847 is reported in the literature to this date.

From atmospheric fits to the observed spectrum and photometry of
W1639$-$6847, \citet{Schneider2015} estimated an effective temperature
of 400 K and a high surface gravity for the target, although such
model-derived physical characteristics are likely to be somewhat
unreliable \citep{Schneider2015}. Based on Gemini spectroscopic data,
\citet{Leggett2017} derived a similar effective temperature (360$-$390
K) as \citet{Schneider2015}, but found a lower surface gravity. Using
evolutionary models, they obtained a mass of 5$-$14 M$_\mathrm{Jup}$
for an age of 0.5$-$5 Gyr. More recently, \citet{Zalesky2019}
performed detailed atmospheric retrieval analyses on late-T and Y
brown dwarfs using \textit{HST} data. While the large majority of
their studied objects appeared consistent with the physics of
radiative-convective equilibrium, the retrieved structure for
W1639$-$6847 was strongly deviating from typical temperature-pressure
profiles under the assumption of radiative-convective equilibrium. The
obtained fit provided rather unrealistic results, with a high
effective temperature of $\sim$650 K, and very small radius (0.5
R$_\mathrm{Jup}$) and mass values (1.5 M$_\mathrm{Jup}$). The authors
concluded that their data-driven atmospheric retrieval was poorly
adapted to explain the deviant physical characteristics of this unique
ultracool brown dwarf.

As noted by several authors, the majority of such analyses are highly
sensitive to the adopted distances of the studied
objects. \citet{Tinney2012} initially derived a parallactic distance
of $5.0\pm0.5$ pc for W1639$-$6847. They also reported a very large
proper motion ($\sim$3 arcsec yr$^{-1}$) and measured a significant
tangential velocity. They deducted from kinematics arguments that the
source was likely older than the overall field population, in
agreement with \citet{Leggett2017} who found it to be consistent with
thin disk membership. \citet{Tinney2014} then refined the proper
motion and parallax estimates, significantly reducing the size of
previous uncertainties.
Recent work by \citet{Martin2018} and \citet{Kirkpatrick2019} provided
updated astrometry for W1639$-$6847 based on \textit{Spitzer} images,
refining its distance to $4.39^{+0.18}_{-0.17}$ pc \citep{Martin2018}
and $4.72\pm0.06$ pc \citep{Kirkpatrick2019}, respectively. Existing
results on parallax based on various datasets remain discrepant by up
to 2.9\,$\sigma$.
Additional and independent reliable astrometric measurements of
W1639$-$6847 will thus be crucial to understand the nature and further
characterise the peculiar features of this distinct object.

\begin{figure*}
\begin{center}
\includegraphics[width=88mm]{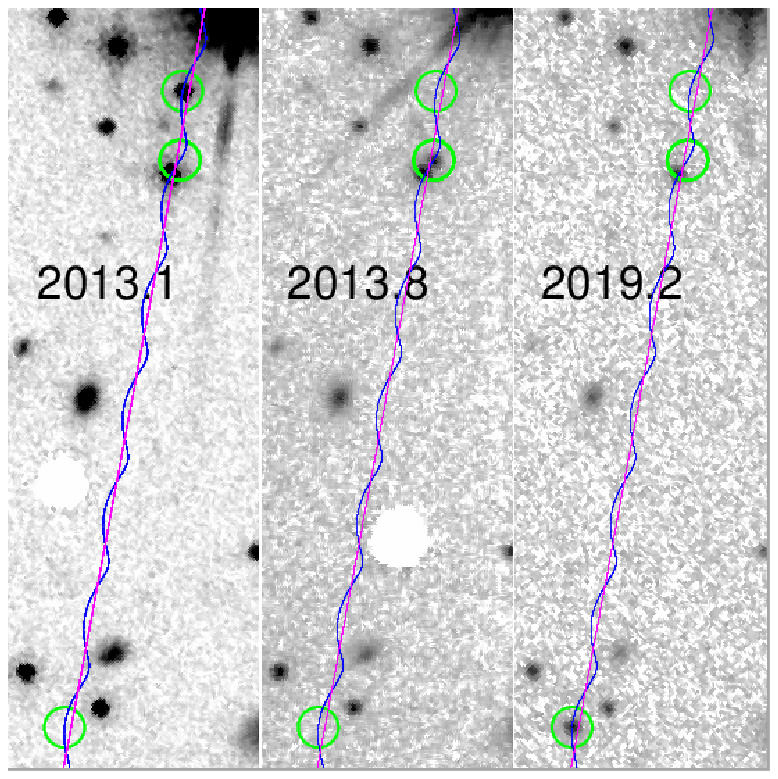}
\includegraphics[width=88mm]{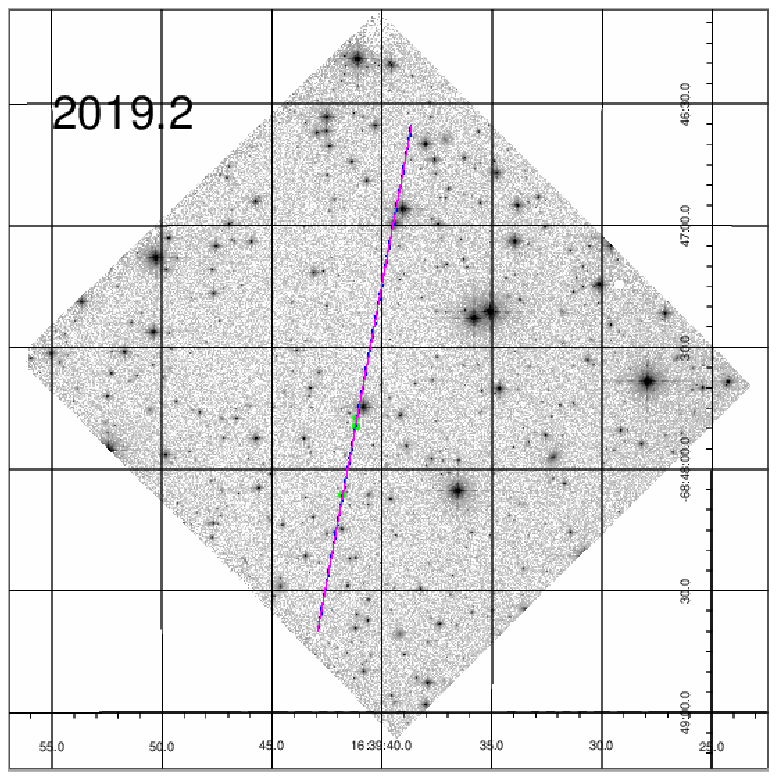} 
\caption{
  \textit{(Left:)} Zoom-in of the \hst\ field surrounding
  W1639$-$6847, as collected in the three main epochs analysed in this
  work.  This small region has a size of $\sim
  7.5^{\prime\prime}\times22.5^{\prime\prime}$.  Green circles
  indicate the BD positions at the three epochs.  Blue and magenta
  lines show our astrometric solution of the motions (see
  Sect.\,\ref{fitAP}) for this object (with and without parallax) in
  years from 1990 to 2030.
   \textit{(Right:)} The entire field of view is about
   2$^{\prime}$$\times$2$^{\prime}$ and this is the stack of four
   WFC3/IR/F127M images collected in last epoch, where the dither
   pattern is best.  The grid and labels are in equatorial
   coordinates.
\label{stack}
}
\end{center}
\end{figure*}

%
\section{Observations} 
\label{observations}
%

W1639$-$6847 was observed at three different epochs with the Wide
Field Camera 3 (WFC3) instrument onboard the \textit{Hubble Space
  Telescope} (\textit{HST}). All data were collected using the
infrared (IR) channel of WFC3. The first epoch was acquired as part of
program GO 12970 (PI: Cushing), on February 15$^{\rm th}$ 2013. The
second epoch consists of three sub-epochs obtained as part of the same
program on October 26$^{\rm th}$, 27$^{\rm th}$ and 29$^{\rm th}$
2013. A final orbit of observations was taken on March 11$^{\rm th}$
2019 for GO 15201 (PI: Fontanive).

The first visit was obtained in the WFC3/IR F125W filter. It was split
into 4 dithered images of 602.937\,s exposure each, for a total
exposure time of 2411.749\,s. Each image was taken in MultiAccum mode
with NSAMP=14 samplings and using the sequence SAMP-SEQ=SPARS50.

The photometric data acquired for W1639$-$6847 on October 26$^{\rm
  th}$ and 27$^{\rm th}$ 2013 each consist of 3 dithered, shallow
images of duration 127.935\,s in the F105W bandpass (SAMP-SEQ=SPARS25
and NSAMP=7) for total exposures of 383.805\,s. The data from October
29$^{\rm th}$ 2013 consist of 4 slightly deeper exposures obtained in
the F125W filter adding up to a combined exposure time of 986.749\,s:
2 images of 277.938\,s each using SAMP-SEQ=SPARS25 and NSAMP=13, and 1
exposure of duration 252.937\,s with the same SAMP-SEQ sequence and
NSAMP=12, and a final image of 177.936\,s with NSAMP=9 samples. The
rest of these orbits were dedicated to spectroscopic observations,
which we do not consider in this work.

The final, most recent epoch consists of one \textit{HST} orbit, split
between the F127M and F139M filters, the combination of these two
bandpasses being highly suited to identify substellar objects through
a deep water absorption feature (see \citealp{Fontanive2018} for
details). In each filter, 4 dithered images of equal duration
(327.939\,s) were acquired in MultiAccum mode, with SAMP-SEQ=SPARS25
and NSAMP=15, for a total exposure time of 1311.756\,s in each
band. Due to the faintness of our target in the F139M filter, only the
F127M data is considered in the astrometric analysis presented in this
work.

Therefore, a total of 18 individual images (4+3+3+4+4) are employed
for the analysis described in the following.

%
\section{Analysis} 
\label{analysis}
%
We first briefly summarise the data reduction and analyses described
in Paper\,I.

We have extracted positions and magnitudes in every single WFC3/IR
\texttt{FLT} image with the software developed by J.\, Anderson
\citep{Anderson&King2006} and publicly available for
WFC3/IR.\footnote{
\texttt{http://www.stsci.edu/$\sim$jayander/WFC3/}
}
This software also produces a quality-of-fit parameter ($Q$;
\citealp{Anderson2008}) that essentially measures how well the flux
distribution resembles the point-spread-functions (PSFs). In these
data sets, the parameter $Q$ is close to 0.02 for the best measured
stars, degrading to $Q\sim0.75$ for the faintest stars. Artefacts,
resolved galaxies, and compromised or blended measurements always have
larger $Q$ values compared to point sources of a same brightness.
The derived positions for detected sources are in raw pixel coordinates and are then corrected for the nominal distortion of the camera, which is also publicly available$^1$.

Given the expected highest signal-to-noise ratio for the sources measured in images from the first epoch, we choose 2013.12427 (Feb.\,15$^{\rm th}$, 2013) as our reference epoch. 
Four images are available for epoch 2013.12427. The distortion-corrected positions for the 
sources measured in all four images are combined to compute 
a more robust estimate of their relative positions. This provides us with 436 sources defining our reference frame $(X,Y)$. 

Next, we link our $(X,Y)$ reference frame to Gaia\,DR2 \citep{GaiaCollaboration2016,GaiaCollaboration2018}, in order to transform our measured positions into the ICRS. 
To do that, Gaia\,DR2 $(\alpha,\delta)$ sources positions, which are given at epoch 2015.5, 
are first re-positioned at the 2013.12427 epoch using (when available) the tabulated proper motions (pms) of those sources. 
Then a tangent point is adopted, and the coordinates on the tangent plane $(\xi,\eta)$ are computed. 
At this point, for all common sources, it becomes possible to compute the most general linear transformations 
to transform any measured position on the master frame into the tangential plane, and then those positions on the 
tangential plane via trivial transformations (see equations 1-4 in Paper\,I) into the ICRS.  
We initially consider all sources in our master frame, including those with no pms in the Gaia\,DR2 catalog.
Once the match is found, we then restrict this sample to sources that are not saturated in the first epoch images, have Gaia\,DR2 pms, and have positions consistent within at most 0.03\,WFC3-pixels (i.e., 3.6\,mas, for the pixel scale 120.9918\,mas derived from this transformation, see Section 3.4 of Paper\,I) between Gaia\,DR2 at the reference epoch and the reference system. This reduced our available number of common sources to 55.

Finally, measured positions in all the images from all epochs can be linked to the very same reference frame $(X,Y)$, now made of Gaia\,DR2 sources that can be re-positioned to the corresponding epoch using the tabulated pms.  
This enables us to transform to the ICRS the positions of every measured object (including sources much fainter than those detectable in Gaia), in every single image, of every single epoch. 
We refer the reader to Paper\,I for a more extensive description of the entire procedure.

\subsection{Improving the method}
In our previous work, when re-positioning the Gaia\,DR2 sources at the corresponding epoch 
of each individual image, we only considered the pms --and not the parallaxes. 
However, the reference sources are all at a finite and different distance, 
which if ignored would inevitably lead to underestimates in the absolute parallax of the target. 
Given the size of the uncertainty ($\sim$2\,mas) in the parallax of
the target of Paper\,I (WISE\,J154151.65$-$225024), and the already
complicated nature of the method, we opted to not add in that work the
further complication of dealing with the individual parallaxes of the
reference sources. Instead, we simply applied an \textit{a posteriori}
correction from relative-to-absolute for the target parallax, which
was of the order of 0.2\,mas (i.e., $<<2$\,mas), and taken as the
median of the Gaia\,DR2 parallaxes of the reference objects (after
rejecting the one with the most significant parallax).

Now that the bulk of the procedure has been presented in Paper\,I, we
further refine our method and develop the procedure to include the
parallaxes of all the reference sources as well.
As we will see, this will turn out to be a rather unnecessary step
given the currently available data for the specific case of
W1639$-$6847 analysed in the present paper.  Nevertheless, it is the
appropriate occasion to improve the method in order to obtain absolute
astrometric parameters, which might be a necessity for future
applications with data sets of higher precisions.\\~

First of all, we need to consider only the sources in Gaia\,DR2 that,
in addition to positions and proper motions, have also a parallax
estimate, and then compute their astrometric place at each of the
observation epochs, this time including their parallaxes.
To compute the positions of the sources in the reference frame, we
make use of the sophisticated tool developed by U.S. Naval
Observatory, the Naval Observatory Vector Astrometry Software,
hereafter \texttt{NOVAS} (in version F3.1, \citealp{Kaplan2011}),
which accounts for many subtle effects, such as the accurate Earth
orbit, perturbations of major bodies, nutation of the Moon-Earth
system, etc.


In particular, we employ the \texttt{NOVAS}'s subroutine
\texttt{ASSTAR}, which computes the astrometric place of a star. This
subroutine takes as input for a source: the ICRS coordinates at epoch
2000.0, the proper motions, the parallax and the radial velocity
(RV, which we set identically to RV=0.0\,km s$^{-1}$
for all sources). The routine in turn produces --at a specified
location in the Solar System-- the astrometric place of the source in
right ascension and declination, at a specified Julian date.
Therefore, as Gaia\,DR2 ICRS positions are given at the epoch 2015.5, we first need to re-position
ICRS coordinates to epoch 2000.0, by using Gaia\,DR2 pms, before passing them to the \texttt{ASSTAR} subroutine.\\~


The need for an existing parallax measurement (and with a positive
value) significantly restricts the sample of usable Gaia\,DR2 sources.
Most of the images have over 20 Gaia\,DR2 sources detected on them
satisfying these criteria, with a maximum of 25 and a minimum of 14
common sources.
Nevertheless, even the image with the minimum number of detected
sources in common with Gaia\,DR2, i.e., 14, has 14$\times$2D positions
that are more than adequate to constrain the six parameters of the
most general linear transformation to bring those detected positions
on that individual image into the ICRS, at the sub-mas level.  [Note
  that for a six parameter transformation, 3$\times$2D data points
  would be sufficient.]  We are thus able to exploit the Gaia\,DR2
reference sources in each of the 18 individual images employed in this
work to carefully study the motion in the field surrounding
W1639$-$6847.

\subsection{Stack Images}
With the coordinate transformations from each image to the reference frame $(X,Y)$ we can create 
stacked images within each epoch, and for each filter. 
Stacked images give the best view of the astronomical scene that can be used to independently 
check the nature of sources in images. 
On left panel of Fig.\ref{stack}, we show the obtained stacks for the three main epochs for 
the two filters with a  similar effective wavelength, i.e., F125W (for 2013.1 and 2013.8) 
and F127M (for 2019.2), in the patch of sky crossed by W1639$-$6847 between these epochs. 
The right panel shows the entire field of view for the F127M observations.  

We saved our stacked images in \texttt{fits} format, and put in their headers our absolute 
astrometric solution with keywords for World Coordinate System. 
These five stacked images --one for each filter/epoch combination-- are released as 
supplementary electronic material of this work. 
Note that the $(X,Y)$ coordinates in this paper are not in the same pixel-coordinate system of these stacked images, 
which are instead super-sampled by a factor two (i.e., each pixel is $\sim$\,60\,mas in size).


\subsection{Determination of the Astrometric Parameters}
\label{fitAP}
Our 18 images (in 2D) gave 36 individual data points, from which to
extract the five astrometric parameters: positions $(X,Y)$, proper
motions $(\mu_X,\mu_Y)$ and parallax ($\varpi$) for W1639$-$6847.  As
motivated in Paper\,I, we keep our calculations in the observational
plane $(X,Y)$.

Again, \texttt{NOVAS} is used to predict the astrometric place of W1639$-$6847.
We then use a Levenberg-Marquardt algorithm (the FORTRAN version
\texttt{lmdif} available under \texttt{MINIPACK}, \citealp{More1980})
to find the minimisation of the [observed$-$calculated] values for the
five parameters: $X,Y,\mu_{X},\mu_{Y},$ and $\varpi$.

\begin{figure*}
\begin{center}
\includegraphics[width=150mm]{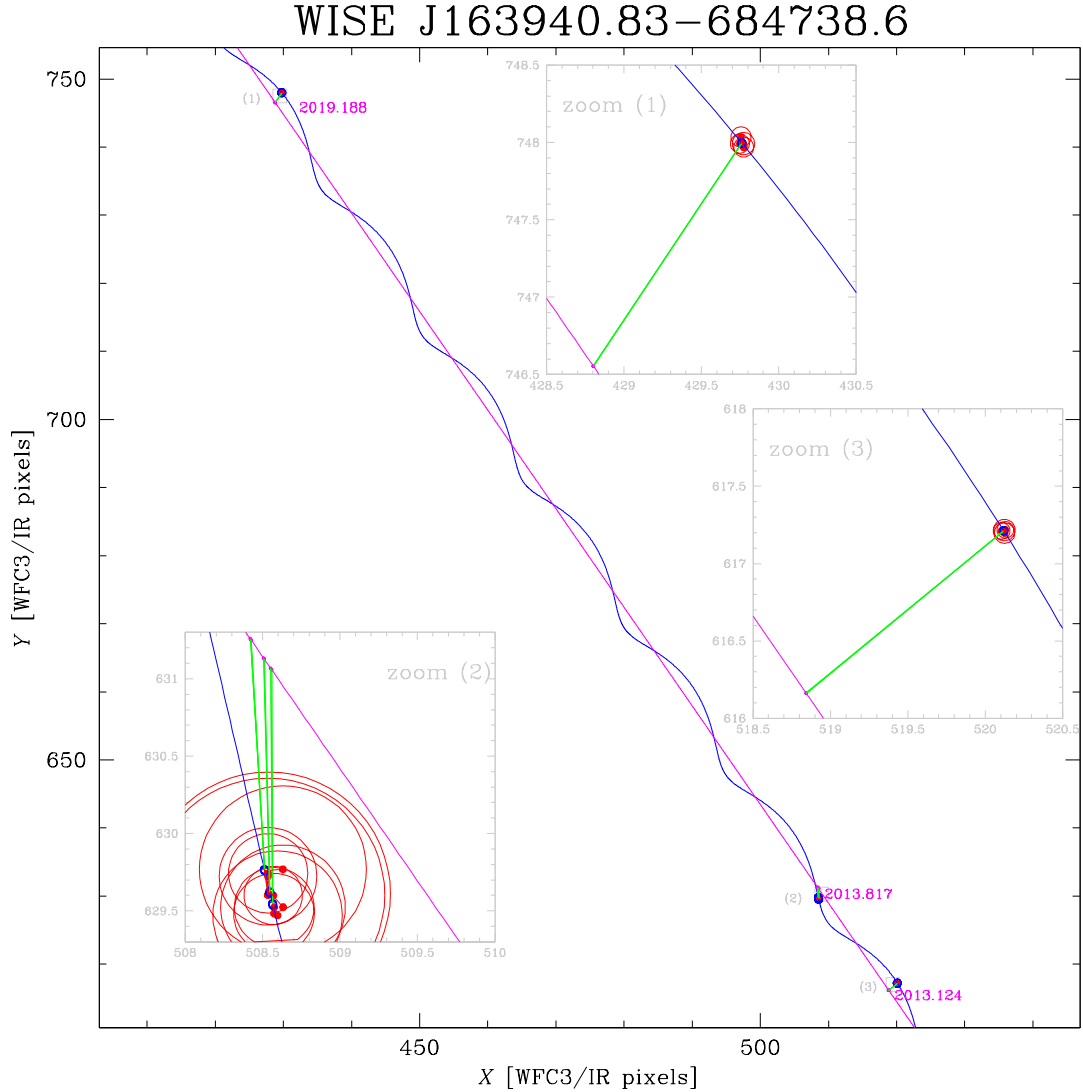}
\caption{
  Comparison of our astrometric solution (line in blue) with the
  individual observed data points (red bullet) for W1639$-$6847 in the
  observational plane $(X,Y)_{2013.12}$. The three major epochs are
  indicated by labels, and three insets indicated by the gray boxes,
  with (1), (2) and (3), show a more meaningful zoom-in of the data
  points.  Red circles indicate the quality fit ($Q$) for each data
  point, with smaller values for better measurements.  To better
  highlight the parallax component of the motion, a line in magenta
  indicates an object with the same motion but at infinite
  distance. Green lines show the parallax contributions at each epoch,
  and red segments connect the individual data points with their
  expected positions (on the blue line) according to the best fit.
\label{cycloid}
}
\end{center}
\end{figure*}

Our best-fit astrometric solution is given in Table\,\ref{tabASTR} and
shown in Fig.\,\ref{cycloid}.
We note that the estimated parallax is already in an absolute reference system. 
To assess the uncertainties of our solution we perform 25\,000
simulations, adding random errors following Gaussian distributions
with dispersion derived from the observed data of W1639$-$6847 for
each of the five filter/epoch combinations (i.e., F125W@2013.1,
F105W@2013.812, F105W@2013.820, F125W@2013.826, and F127M@2019.2).
The intrinsic $\sim$0.050\,mas systematic uncertainties inherent to
the Gaia\,DR2 parallaxes \citep{Lindegren18} need to be added to the
error budget, although completely insignificant compared to the
estimated errors on the parallax.

We note that the two epochs with the widest time-baseline also have
the best astrometric accuracies and are taken almost exactly at the
same phase of the year, making our pms exquisitely accurate, at a
quart-of-a-mas level.  However, with only three phases of the year
mapped, the parallax estimate relies entirely on (and is therefore
limited by) the weakest measurements at epochs $\sim$2013.8.

The astrometric precisions at this epoch are, unfortunately,
significantly worse than those at the other two epochs for several
reasons.
First, all images within this epoch are affected by contaminating
light coming primarily from scattered Earthlight. This anomaly is
often present for IR observations made when the limb angle, which is
the angle between \textit{HST}'s pointing direction and the nearest
limb of the bright Earth, is less than $\sim$30 degrees.
Second, the total exposure times, and therefore the average signal to
noise ratios in each of these images are significantly lower than for
those taken in 2013.1 and 2019.2.
Third, the close proximity of a relatively bright star at $\sim$3.5
pixels from W1639$-$6847 might also have contributed to enlarge the
errors (see Fig.\ref{stack}).\\~

\begin{figure}
\begin{center}
\includegraphics[width=88mm]{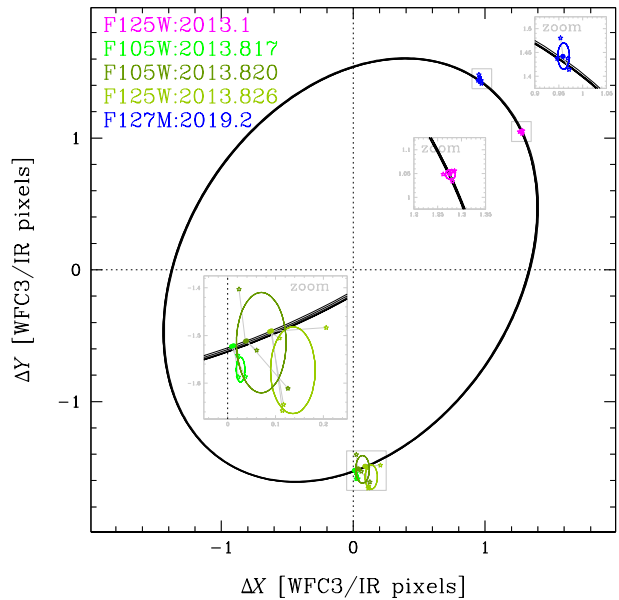}
\caption{
  Our solution for the parallax ellipse in the $(X,Y)_{2013.12}$
  coordinate system.  Individual \textit{HST} data points are
  indicated with star symbols, which are connected with small segments
  to their expected positions according to our best fit.  Smaller
  ellipses in magenta, green, and blue, indicate the 1-$\sigma_{X,Y}$
  of individual data points within each epoch. 
  Note how ellipses are significantly smaller for the first and last epochs, 
  compared to the 2013.8 sub-epochs. Insets in
  gray have the same scale, and show zoom-in views around the locations 
  marked by gray boxes.
\label{Ell}
}
\end{center}
\end{figure}

In addition to Fig.\,\ref{cycloid} and its insets, we show in
Fig.\,\ref{Ell} the parallax ellipse along with \textit{HST}
measurements \textit{[proper motion subtracted].}  This representation
better reveals the sampling of the parallactic motion which, with only
three main epochs, could be problematic.
The fact that the 2013.8 epoch is made of three sub-epochs separated
by about a day (on 26, 27 and 29 October 2013, respectively) slightly
alleviates this problematic situation in the parallax estimate.
While our parallax best fit provides a \textit{formal} error of only
$\sim$2\,mas, a close look at our best-fit compared to the observed
points at these $\sim$2013.8 epochs seems to suggest a marginally
larger parallax, which could be larger by as much as
$\sim$0.04\,WFC3/IR pixel (i.e., $\sim$5\,mas), or possibly residuals
caused by the closeness to the aforementioned field-star at
$\sim$3.5\,pixels in that epoch.
Indeed, with only 3 main annual phases probed, it is hard to highlight
the presence of unaccounted systematic errors in these values.
A single future measurement could be sufficient to significantly
refine and consolidate our new parallax estimate.

%
\begin{table}
\caption{Astrometric parameters of W1639$-$6847 in the
      ICRS. Positions are given at 2000.0 and at the Gaia\,DR2 2015.5 epoch,
      where the $^\varpi$ suffix indicates that the apparent positions have
      the annual parallax included.}
\center
\begin{tabular}{lcc}
\hline
%
& & \\
$\alpha_{2000.0}$ [ $^{\rm h}$ $^{\rm m}$ $^{\rm s}$ ]         &    16:39:39.72931  & $\pm$ 11\,mas \\
$\delta_{2000.0}$ [ $^{\circ}$ $^{\prime}$ $^{\prime\prime}$ ] & $-$68:47:06.69404  & $\pm$ 5\,mas \\
& & \\
$\alpha_{2000.0}$ [degrees] &   249.9155388  & $\pm$ 11\,mas \\
$\delta_{2000.0}$ [degrees] & $-$68.78519279 & $\pm$ 5\,mas \\
& & \\
$\alpha_{2015.5}$ [degrees] &   249.9224084 & $\pm$ 4.5  \,mas \\ 
$\delta_{2015.5}$ [degrees] & $-$68.79857156 & $\pm$ 0.8  \,mas \\ 
& & \\
$\alpha_{2015.5}^\varpi$ [degrees] &   249.9223335  & $\pm$  6.6  \,mas \\ 
$\delta_{2015.5}^\varpi$ [degrees] & $-$68.79860735 & $\pm$  1.5  \,mas \\ 
& & \\
$\mu_{\alpha\cos{\delta}}$ [mas yr$^{-1}$]  &  $+$577.21 & $\pm$ 0.24 \\
$\mu_{\delta}$ [mas yr$^{-1}$]              & $-$3108.39 & $\pm$ 0.27 \\
& & \\
%
$\varpi$ [mas]              & 210.35     & $\pm$ 1.82 $\pm$0.05 \\
%
\hline
\end{tabular}
\label{tabASTR}
\end{table} 
%

%
\begin{table*}
  \caption{
    List of works in the Literature providing astrometric parameters for W1639$-$6847.
    }
  \center
\begin{tabular}{lccccr}
\hline
work  & $\mu_{\alpha^*}\pm\sigma\mu_{\alpha^*}$ & $\mu_\delta\pm\sigma\mu_\delta$ & $\varpi\pm\sigma_\varpi$ & $d$ & source\\
\#. authors (date) & [mas\,yr$^{-1}$] & [mas\,yr$^{-1}$] & [mas] & [pc] & facilities \\
\hline
   1. \citet{Tinney2012}     & $580\pm22$    & $-3013\pm40$      & $200\pm20$    & $5.0\pm0.5$   & Magellan+\textit{WISE} \\
   2. \citet{Tinney2014}     & $586.0\pm5.5$ & $-3101.1\pm3.6$   & $202.3\pm3.1$ & $4.9\pm0.1$   & Magellan \\
   3. \citet{Pinfield2014}   & $-800\pm1200$ & $-2800\pm1200$    & ...           & ...           & \textit{WISE} \\
   4. \citet{Martin2018}     & $579.09\pm12.52$ & $-3104.54\pm12.25$ & $228.05\pm8.93$ & $4.39^{+0.18}_{-0.17}$ & \textit{Spitzer} \\
   5. \citet{Kirkpatrick2019}& $582.0\pm1.5$ & $-3099.8\pm1.5$ & $211.9\pm2.7$ & $4.72\pm0.06$ & \textit{Spitzer} \\
   6. \textbf{this work}     & $\mathbf{577.21\pm0.24}$ & $\mathbf{-3108.39\pm0.27}$  & $\mathbf{210.35\pm1.82}$ & $\mathbf{4.75\pm0.05}$   & \textit{\textbf{HST+Gaia}} \\
\hline
\end{tabular}
\label{lit}
\end{table*} 
%

\subsection{Improved- \textit{vs.\,} old-method, and RVs}
%
Even if we expect negligible differences for the case of W1639$-$6847, 
it is worth comparing the numerical results of our procedure from Paper\,I with the 
new procedure presented in this work, which includes the parallaxes for the Gaia\,DR2 reference sources. 

\textit{In our first test}, we performed the astrometric parameters fit using the very same sample of 
reference stars in each image (14--25), but this time not including their parallaxes 
(i.e., assuming them to be at infinite distances, therefore setting their parallaxes to zero). 
We obtained a parallax of $\pi$=209.74\,mas, which is slightly smaller than 
the value $\varpi$=210.35\,mas obtained in Table\,\ref{tabASTR}. 
This reduced parallax for W1639$-$6847 goes in the right direction,
meaning that it is an apparent parallax ($\pi$) which is obtained with
respect to reference sources that are not at infinite distances.
Therefore, $\pi$ is smaller than the absolute parallax ($\varpi$), as
it does not contain the parallax of the references sources, hence
expected to be smaller.
However, it is only a marginally smaller value, as $\sim$0.6\,mas
compared with an estimated uncertainty for $\varpi$ of 1.8\,mas
(1-$\sigma$) corresponds to a $\sim$0.3\,$\sigma$ significance.
Finally, we note that all the other astrometric parameters (positions,
and proper motions) show even less significant changes.
\textit{As a second test}, we compute transformations using all the
Gaia\,DR2 stars with proper motions, even when no (positive)
parallaxes were available. This results in an enlarged sample of
reference objects (57-79 vs. 14-25).
The derived apparent parallax in this case is $\pi$=210.02\,mas, thus
even closer ($\sim$0.3\,mas) to our derived absolute parallax
($\varpi$=210.35\,mas), and consistent with it at the
$\sim$0.2\,$\sigma$-level.
We note that the consistency in positions between Gaia\,DR2 stars and
their positions in the \textit{HST} images are always better than
$\sim$3\,mas (Paper\,I, Fig.\,3, as well as this work), and that the
inconsistencies are dominated by random errors in the positions
measured in the \textit{HST} images. Therefore, going from $\sim$20 to
$\sim$70 reference sources we could hope to reduce the errors in our
transformations (from the image coordinates system of the \textit{HST}
individual images to the Gaia\,DR2 system) at most from
$\sim$0.65\,mas to $\sim$0.35\,mas, which are both well within the
uncertainties of our individual measurements, and also within the
errors in our fitted absolute parallax, $\sigma_\varpi$=1.8\,mas.

\textit{In our third and last test}, we explore the impact of RVs on
our final astrometry.
In our derivations of the astrometric parameters, we have assumed the
RVs for all the stars, W1639$-$6847 included, to be identically zero.
However, a non-null radial velocity means that objects change in time
their distance with respect to the observer, and therefore change
their parallax in time, as result of projection effects.
Essentially all reference stars in our studied field are significantly
further away than our science target. Therefore neglecting their RVs
has a much smaller effects than neglecting the target RV, as their
distances will change by much less in percentage than for
W1639$-$6847.
Assuming arbitrary RV values for the target simply cause fluctuations
of our fit within the noise, for RVs up to $\pm$1000\,km\,s$^{-1}$.
This is not surprising, as even for the most extreme case of the
\textit{Barnard's Runaway Star}, which has an RV of
$-$110.6\,km\,s$^{-1}$ and a $\varpi$=547.45\,mas, we expect a
parallax change rate of only $\dot{\varpi}=+34$\,$\mu$as\,\,yr$^{-1}$
\citep{dravins1999}.

Nevertheless, it is interesting to note that astrometry could be used,
\textit{in turn}, to estimate RVs, and that these astrometric-RVs do
not suffer from spectroscopic biases such as gravitational redshifts
(as high as 25\,km\,s$^{-1}$ for WDs), convective bubble motions
($\sim$0.5\,km\,s$^{-1}$ for red giants), etc. (indeed, any
spectroscopic measurement is always model-dependent, while astrometry
is a purely geometrical one).
The secular changes of trigonometric parallaxes are well known effects
that can be used to determine model-independent astrometric-RVs (see
paper series by \citealp{dravins1999} for a review).  Astrometric-RVs
are well within the reach of \textit{Gaia} precision for several
close-by (or fast-moving) stars, but extremely hard to measure with
traditional \textit{HST} images (at least in non-trailing mode).

%
\section{Conclusions}
%

In this work, we have perfected the procedure developed in Paper\,I
\citep{Bedin&Fontanive2018} exploiting the power of Gaia DR2 to
improve imaging-astrometry with narrow-field cameras. Our method makes
use of the positions, proper motions and parallaxes of stars in the
Gaia DR2 catalog to derive highly-precise astrometric solutions for
sources too faint for Gaia observed in multiple epochs of \textit{HST}
data. The technique was refined in this paper to include the Gaia DR2
parallaxes of the astrometric reference sources in the analysis,
allowing us to directly obtain absolute astrometric parameters.

This improved procedure was applied to the brown dwarf
WISE\,J163940.83$-$684738.6, a Y0pec dwarf with puzzling photometric
and spectroscopic features. The distance and proper motion of this
unusual object are poorly constrained, with significant
inconsistencies between existing estimates. Using three epochs of
\textit{HST}/WFC3 data acquired over a period of $\sim$6 years, we
were able to constrain its parallax to $\varpi=210.4\pm1.8$\,mas, and
its proper motion to
$\mu_{\alpha\cos{\delta}}=577.21\pm0.24$\,mas\,yr$^{-1}$,
$\mu_\delta=-3108.39\pm0.27$\,mas\,yr$^{-1}$.

With achieved precisions of $\sim$2\,mas in parallax and at the
sub-mas level in proper motion, these new astrometric parameters
represent considerable improvements relative to previous estimates, as
summarised in Table~\ref{lit}. On one hand, our proper motion
measurements are in good agreement with other estimates from the
literature. In particular, our derived $\mu_{\alpha\cos{\delta}}$ and
$\mu_\delta$ values are consistent with the results from
\citet{Tinney2014} and \citet{Martin2018} within 2\,$\sigma$, although
our obtained uncertainties are smaller by more than an order of
magnitude.
On the other hand, larger disparities ($>$3\,$\sigma$) are observed
between our proper motion measurements and those from
\citet{Kirkpatrick2019}, which were the most accurate to date.

Our estimates of the astrometric parameters for W1639$-$6847 are
completely independent from the ones obtained with \textit{Spitzer}
data, and because of this, have an important value on their own.  For
the same reason it would also be interesting to combine them properly.
Indeed, while unaccounted systematic errors in our estimated parallax
could be as large as $\sim$5\,mas, due to the problematic epochs
around 2013.8 (see Sect\,4.3), based on our experience we can hardly
expect residual systematic errors larger than 1\,mas\,yr$^{-1}$ in the
estimated proper motions derived from \textit{HST} data (e.g.,
\citealt{Bellini2018} and reference therein).
As we do not have the competence to analyse \textit{Spitzer} data at
the same level of accuracy as we have done for the \textit{HST} data
(not only distortion and positioning, but particularly the way to
simultaneously fit \textit{HST} data with data from a telescope in a
significantly different, Earth-trailing, Heliocentric orbit), we list
in Table\,\ref{positions} our \textit{HST} individual measurements to
allow future investigators to be able to properly combine the two
space-based datasets.\\ 
%

In terms of parallax, results from previous works were more
discordant, with a $\sim$10\% discrepancy between the best available
estimates so far
\citep{Tinney2014,Martin2018,Kirkpatrick2019}. Interestingly, our
newly-derived value was found to be somewhere in between the
ground-based and \textit{Spitzer} determinations from
\citet{Tinney2014} and \citet{Martin2018}, respectively, and this time
in excellent agreement with the \textit{Spitzer}-derived value from
\citet{Kirkpatrick2019}, which used additional epochs of data compared
to the work from \citet{Martin2018}. The corresponding distance of
$4.75\pm0.05$\,pc we obtained here for W1639$-$6847 makes our result
the most accurate distance measurement available for this Y dwarf.

As previously discussed, our parallax estimate for W1639$-$6847 relies
entirely on the epoch with the lowest astrometric precision, and will
require an additional epoch of observations to be further validated
and refined. An accurate measurement of the distance to W1639$-$6847
will certainly be the key to modelling and understanding the peculiar
atmospheric characteristics observed to date for this object.
Nevertheless, we have successfully demonstrated that our powerful
procedure allows us to place strong constraints on the parallax and
proper motion of extremely faint objects, based on only three epochs
of observations taken over a baseline of $\sim$half a decade.

The \textit{Hubble Space Telescope} indeed provides a unique
opportunity to reach such results for faint and red brown dwarfs, with
an ideal comprise between the $\sim$121\,mas plate scale of the
WFC3/IR channel and the wide field of view allowing for numerous
astrometric references, combined with the exquisite stability achieved
from space.  In contrast, other space-based telescopes generally have
significantly broader pixel sizes ($>$1$-$2 arcsec), leading to lower
astrometric resolutions and increased chances of blended sources (like
it was originally the case for our target W1639$-$6847 in
\textit{WISE}; \citealp{Tinney2012}).  While ground-based facilities
typically have much higher angular resolutions, mitigating the broad
plate scale drawbacks, observations from the ground are constrained by
sensitivity, rending observations of the faintest brown dwarfs
extremely challenging. In addition, ground-based data generally suffer
from atmospheric aberrations and numerous systematic errors that can
be difficult to quantify and account for when comparing between
near-infrared brown dwarf targets and field stars of very different
colours.

\textit{HST} therefore represents a superior platform for
high-precision astrometry of ultracool dwarfs, and for a method like
the one developed in this paper to be applied. The derivation of new
distance measurements for a number of additional Y brown dwarfs via
such an approach will be crucial to the characterisation of these
objects, and will undoubtedly shed new light on substellar studies, at
the individual and population levels.

The remarkable spatial and spectral resolution of the anticipated
\textit{James Webb Space Telescope} (\textit{JWST}) will soon allow
for unparalleled probes of ultracool brown dwarfs at near-infrared
wavelengths, by observing at wavelengths where Y dwarfs are orders of
magnitude brighter than they are at \textit{HST} wavelengths.
In particular, between the very large field of view of the Near
Infrared Camera (NIRCam) instrument and its exceptional angular
resolution of 32\,mas at 2\,$\mu$m, we will be able to take our
technique a step further with \textit{JWST}, and measure precise
distances to the coldest objects in the Solar neighbourhood to even
greater accuracies. This will in turn tremendously enhance our
understanding of planet-like atmospheres and will provide unique
opportunities to calibrate theoretical models at the low-mass end of
the substellar regime.\\~

\section*{Acknowledgements}
We thank an anonymous Referee for the useful comments, the tests
suggested, and for the prompt review of our work.
This work is based on observations with the NASA/ESA Hubble Space
Telescope, obtained at the Space Telescope Science Institute, which is
operated by AURA, Inc., under NASA contract NAS 5-26555.
This work makes also use of results from the European Space Agency
(ESA) space mission Gaia. Gaia data are being processed by the Gaia
Data Processing and Analysis Consortium (DPAC). Funding for the DPAC
is provided by national institutions, in particular the institutions
participating in the Gaia MultiLateral Agreement (MLA). The Gaia
mission website is \texttt{https://www.cosmos.esa.int/gaia}. The Gaia
archive website is \texttt{https://archives.esac.esa.int/gaia}.
This research has benefitted from the Y Dwarf Compendium maintained by
Michael Cushing at \texttt{https://sites.google.com/view/ydwarfcompendium/}.
LRB acknowledges support by MIUR under PRIN program \#2017Z2HSMF.


\bibliographystyle{mnras}
\bibliography{biblio}

\begin{thebibliography}{}
\makeatletter
\relax
\def\mn@urlcharsother{\let\do\@makeother \do\$\do\&\do\#\do\^\do\_\do\%\do\~}
\def\mn@doi{\begingroup\mn@urlcharsother \@ifnextchar [ {\mn@doi@}
  {\mn@doi@[]}}
\def\mn@doi@[#1]#2{\def\@tempa{#1}\ifx\@tempa\@empty \href
  {http://dx.doi.org/#2} {doi:#2}\else \href {http://dx.doi.org/#2} {#1}\fi
  \endgroup}
\def\mn@eprint#1#2{\mn@eprint@#1:#2::\@nil}
\def\mn@eprint@arXiv#1{\href {http://arxiv.org/abs/#1} {{\tt arXiv:#1}}}
\def\mn@eprint@dblp#1{\href {http://dblp.uni-trier.de/rec/bibtex/#1.xml}
  {dblp:#1}}
\def\mn@eprint@#1:#2:#3:#4\@nil{\def\@tempa {#1}\def\@tempb {#2}\def\@tempc
  {#3}\ifx \@tempc \@empty \let \@tempc \@tempb \let \@tempb \@tempa \fi \ifx
  \@tempb \@empty \def\@tempb {arXiv}\fi \@ifundefined
  {mn@eprint@\@tempb}{\@tempb:\@tempc}{\expandafter \expandafter \csname
  mn@eprint@\@tempb\endcsname \expandafter{\@tempc}}}

\bibitem[\protect\citeauthoryear{{Allen}, {Koerner}, {Reid}  \&
  {Trilling}}{{Allen} et~al.}{2005}]{Allen2005}
{Allen} P.~R.,  {Koerner} D.~W.,  {Reid} I.~N.,   {Trilling} D.~E.,  2005,
  \mn@doi [\apj] {10.1086/429548}, \href
  {http://adsabs.harvard.edu/abs/2005ApJ...625..385A} {625, 385}

\bibitem[\protect\citeauthoryear{{Anderson} \& {King}}{{Anderson} \&
  {King}}{2006}]{Anderson&King2006}
{Anderson} J.,  {King} I.~R.,  2006, Technical report, {PSFs, Photometry, and
  Astronomy for the ACS/WFC}

\bibitem[\protect\citeauthoryear{{Anderson} et~al.,}{{Anderson}
  et~al.}{2008}]{Anderson2008}
{Anderson} J.,  et~al., 2008, \mn@doi [\aj] {10.1088/0004-6256/135/6/2055},
  \href {https://ui.adsabs.harvard.edu/abs/2008AJ....135.2055A} {135, 2055}

\bibitem[\protect\citeauthoryear{{Bedin} \& {Fontanive}}{{Bedin} \&
  {Fontanive}}{2018}]{Bedin&Fontanive2018}
{Bedin} L.~R.,  {Fontanive} C.,  2018, \mn@doi [\mnras]
  {10.1093/mnras/sty2626}, \href
  {http://adsabs.harvard.edu/abs/2018MNRAS.481.5339B} {481, 5339}

\bibitem[\protect\citeauthoryear{{Beichman}, {Gelino}, {Kirkpatrick},
  {Cushing}, {Dodson-Robinson}, {Marley}, {Morley}  \& {Wright}}{{Beichman}
  et~al.}{2014}]{Beichman2014}
{Beichman} C.,  {Gelino} C.~R.,  {Kirkpatrick} J.~D.,  {Cushing} M.~C.,
  {Dodson-Robinson} S.,  {Marley} M.~S.,  {Morley} C.~V.,   {Wright} E.~L.,
  2014, \mn@doi [\apj] {10.1088/0004-637X/783/2/68}, \href
  {http://adsabs.harvard.edu/abs/2014ApJ...783...68B} {783, 68}

\bibitem[\protect\citeauthoryear{{Bellini} et~al.,}{{Bellini}
  et~al.}{2018}]{Bellini2018}
{Bellini} A.,  et~al., 2018, \mn@doi [\apj] {10.3847/1538-4357/aaa3ec}, \href
  {https://ui.adsabs.harvard.edu/abs/2018ApJ...853...86B} {853, 86}

\bibitem[\protect\citeauthoryear{{Burgasser}}{{Burgasser}}{2004}]{Burgasser2004}
{Burgasser} A.~J.,  2004, \mn@doi [\apjs] {10.1086/424386}, \href
  {http://adsabs.harvard.edu/abs/2004ApJS..155..191B} {155, 191}

\bibitem[\protect\citeauthoryear{{Chiu}, {Fan}, {Leggett}, {Golimowski},
  {Zheng}, {Geballe}, {Schneider}  \& {Brinkmann}}{{Chiu}
  et~al.}{2006}]{Chiu2006}
{Chiu} K.,  {Fan} X.,  {Leggett} S.~K.,  {Golimowski} D.~A.,  {Zheng} W.,
  {Geballe} T.~R.,  {Schneider} D.~P.,   {Brinkmann} J.,  2006, \mn@doi [\aj]
  {10.1086/501431}, \href {http://adsabs.harvard.edu/abs/2006AJ....131.2722C}
  {131, 2722}

\bibitem[\protect\citeauthoryear{{Cruz} et~al.,}{{Cruz}
  et~al.}{2007}]{Cruz2007}
{Cruz} K.~L.,  et~al., 2007, \mn@doi [\aj] {10.1086/510132}, \href
  {http://adsabs.harvard.edu/abs/2007AJ....133..439C} {133, 439}

\bibitem[\protect\citeauthoryear{{Cruz}, {Kirkpatrick}  \& {Burgasser}}{{Cruz}
  et~al.}{2009}]{Cruz2009}
{Cruz} K.~L.,  {Kirkpatrick} J.~D.,   {Burgasser} A.~J.,  2009, \mn@doi [\aj]
  {10.1088/0004-6256/137/2/3345}, \href
  {http://adsabs.harvard.edu/abs/2009AJ....137.3345C} {137, 3345}

\bibitem[\protect\citeauthoryear{{Cushing} et~al.,}{{Cushing}
  et~al.}{2011}]{Cushing2011}
{Cushing} M.~C.,  et~al., 2011, \mn@doi [\apj] {10.1088/0004-637X/743/1/50},
  \href {http://adsabs.harvard.edu/abs/2011ApJ...743...50C} {743, 50}

\bibitem[\protect\citeauthoryear{{Dravins}, {Lindegren}  \& {Madsen}}{{Dravins}
  et~al.}{1999}]{dravins1999}
{Dravins} D.,  {Lindegren} L.,   {Madsen} S.,  1999, \aap, \href
  {https://ui.adsabs.harvard.edu/abs/1999A&A...348.1040D} {348, 1040}

\bibitem[\protect\citeauthoryear{{Dupuy} \& {Kraus}}{{Dupuy} \&
  {Kraus}}{2013}]{Dupuy2013}
{Dupuy} T.~J.,  {Kraus} A.~L.,  2013, \mn@doi [Science]
  {10.1126/science.1241917}, \href
  {http://adsabs.harvard.edu/abs/2013Sci...341.1492D} {341, 1492}

\bibitem[\protect\citeauthoryear{{Dupuy} \& {Liu}}{{Dupuy} \&
  {Liu}}{2012}]{Dupuy2012}
{Dupuy} T.~J.,  {Liu} M.~C.,  2012, \mn@doi [\apjs]
  {10.1088/0067-0049/201/2/19}, \href
  {http://adsabs.harvard.edu/abs/2012ApJS..201...19D} {201, 19}

\bibitem[\protect\citeauthoryear{{Faherty} et~al.,}{{Faherty}
  et~al.}{2012}]{Faherty2012}
{Faherty} J.~K.,  et~al., 2012, \mn@doi [\apj] {10.1088/0004-637X/752/1/56},
  \href {http://adsabs.harvard.edu/abs/2012ApJ...752...56F} {752, 56}

\bibitem[\protect\citeauthoryear{{Fontanive}, {Biller}, {Bonavita}  \&
  {Allers}}{{Fontanive} et~al.}{2018}]{Fontanive2018}
{Fontanive} C.,  {Biller} B.,  {Bonavita} M.,   {Allers} K.,  2018, \mn@doi
  [\mnras] {10.1093/mnras/sty1682}, \href
  {https://ui.adsabs.harvard.edu/abs/2018MNRAS.479.2702F} {479, 2702}

\bibitem[\protect\citeauthoryear{{Gaia Collaboration} et~al.,}{{Gaia
  Collaboration} et~al.}{2016}]{GaiaCollaboration2016}
{Gaia Collaboration} et~al., 2016, \mn@doi [\aap]
  {10.1051/0004-6361/201629272}, \href
  {https://ui.adsabs.harvard.edu/abs/2016A&A...595A...1G} {595, A1}

\bibitem[\protect\citeauthoryear{{Gaia Collaboration} et~al.,}{{Gaia
  Collaboration} et~al.}{2018}]{GaiaCollaboration2018}
{Gaia Collaboration} et~al., 2018, \mn@doi [\aap]
  {10.1051/0004-6361/201833051}, \href
  {https://ui.adsabs.harvard.edu/abs/2018A&A...616A...1G} {616, A1}

\bibitem[\protect\citeauthoryear{{Kaplan}, {Bartlett}, {Monet}, {Bangert}  \&
  {Puatua}}{{Kaplan} et~al.}{2011}]{Kaplan2011}
{Kaplan} G.,  {Bartlett} J.,  {Monet} A.,  {Bangert} J.,   {Puatua} W.,  2011.
  User's Guide to NOVAS Version F3.1. USNO, Washington, DC

\bibitem[\protect\citeauthoryear{{Kirkpatrick} et~al.,}{{Kirkpatrick}
  et~al.}{2011}]{Kirkpatrick2011}
{Kirkpatrick} J.~D.,  et~al., 2011, \mn@doi [\apj]
  {10.1088/0067-0049/197/2/19}, \href
  {http://adsabs.harvard.edu/abs/2011ApJS..197...19K} {197, 19}

\bibitem[\protect\citeauthoryear{{Kirkpatrick} et~al.,}{{Kirkpatrick}
  et~al.}{2012}]{Kirkpatrick2012}
{Kirkpatrick} J.~D.,  et~al., 2012, \mn@doi [\apj]
  {10.1088/0004-637X/753/2/156}, \href
  {http://adsabs.harvard.edu/abs/2012ApJ...753..156K} {753, 156}

\bibitem[\protect\citeauthoryear{{Kirkpatrick} et~al.,}{{Kirkpatrick}
  et~al.}{2019}]{Kirkpatrick2019}
{Kirkpatrick} J.~D.,  et~al., 2019, \mn@doi [\apjs] {10.3847/1538-4365/aaf6af},
  \href {http://adsabs.harvard.edu/abs/2019ApJS..240...19K} {240, 19}

\bibitem[\protect\citeauthoryear{{Knapp} et~al.,}{{Knapp}
  et~al.}{2004}]{Knapp2004}
{Knapp} G.~R.,  et~al., 2004, \mn@doi [\aj] {10.1086/420707}, \href
  {http://adsabs.harvard.edu/abs/2004AJ....127.3553K} {127, 3553}

\bibitem[\protect\citeauthoryear{{Leggett}, {Tremblin}, {Esplin}, {Luhman}  \&
  {Morley}}{{Leggett} et~al.}{2017}]{Leggett2017}
{Leggett} S.~K.,  {Tremblin} P.,  {Esplin} T.~L.,  {Luhman} K.~L.,   {Morley}
  C.~V.,  2017, \mn@doi [\apj] {10.3847/1538-4357/aa6fb5}, \href
  {http://adsabs.harvard.edu/abs/2017ApJ...842..118L} {842, 118}

\bibitem[\protect\citeauthoryear{{Lindegren} et~al.,}{{Lindegren}
  et~al.}{2018}]{Lindegren18}
{Lindegren} L.,  et~al., 2018, \mn@doi [\aap] {10.1051/0004-6361/201832727},
  \href {https://ui.adsabs.harvard.edu/abs/2018A&A...616A...2L} {616, A2}

\bibitem[\protect\citeauthoryear{{Liu} et~al.,}{{Liu} et~al.}{2011}]{Liu2011}
{Liu} M.~C.,  et~al., 2011, \mn@doi [\apj] {10.1088/0004-637X/740/2/108}, \href
  {http://adsabs.harvard.edu/abs/2011ApJ...740..108L} {740, 108}

\bibitem[\protect\citeauthoryear{{Liu}, {Dupuy}  \& {Allers}}{{Liu}
  et~al.}{2016}]{Liu2016}
{Liu} M.~C.,  {Dupuy} T.~J.,   {Allers} K.~N.,  2016, \mn@doi [\apj]
  {10.3847/1538-4357/833/1/96}, \href
  {http://adsabs.harvard.edu/abs/2016ApJ...833...96L} {833, 96}

\bibitem[\protect\citeauthoryear{{Manjavacas}, {Goldman}, {Reffert}  \&
  {Henning}}{{Manjavacas} et~al.}{2013}]{Manjavacas2013}
{Manjavacas} E.,  {Goldman} B.,  {Reffert} S.,   {Henning} T.,  2013, \mn@doi
  [\aap] {10.1051/0004-6361/201321720}, \href
  {https://ui.adsabs.harvard.edu/abs/2013A&A...560A..52M} {560, A52}

\bibitem[\protect\citeauthoryear{{Manjavacas} et~al.,}{{Manjavacas}
  et~al.}{2019}]{Manjavacas2019}
{Manjavacas} E.,  et~al., 2019, \mn@doi [\aj] {10.3847/1538-3881/aaf88f}, \href
  {http://adsabs.harvard.edu/abs/2019AJ....157..101M} {157, 101}

\bibitem[\protect\citeauthoryear{{Martin} et~al.,}{{Martin}
  et~al.}{2018}]{Martin2018}
{Martin} E.~C.,  et~al., 2018, \mn@doi [\apj] {10.3847/1538-4357/aae1af}, \href
  {http://adsabs.harvard.edu/abs/2018ApJ...867..109M} {867, 109}

\bibitem[\protect\citeauthoryear{Mor\'e, Garbow  \& Hillstrom}{Mor\'e
  et~al.}{1980}]{More1980}
Mor\'e J.~J.,  Garbow B.~S.,   Hillstrom K.~E.,  1980. User Guide for
  MINPACK-1. Argonne National Laboratory Report ANL-80-74, Argonne, IL

\bibitem[\protect\citeauthoryear{{Opitz}, {Tinney}, {Faherty}, {Sweet},
  {Gelino}  \& {Kirkpatrick}}{{Opitz} et~al.}{2016}]{Opitz2016}
{Opitz} D.,  {Tinney} C.~G.,  {Faherty} J.~K.,  {Sweet} S.,  {Gelino} C.~R.,
  {Kirkpatrick} J.~D.,  2016, \mn@doi [\apj] {10.3847/0004-637X/819/1/17},
  \href {https://ui.adsabs.harvard.edu/abs/2016ApJ...819...17O} {819, 17}

\bibitem[\protect\citeauthoryear{{Pinfield}, {Jones}, {Lucas}, {Kendall},
  {Folkes}, {Day-Jones}, {Chappelle}  \& {Steele}}{{Pinfield}
  et~al.}{2006}]{Pinfield2006}
{Pinfield} D.~J.,  {Jones} H.~R.~A.,  {Lucas} P.~W.,  {Kendall} T.~R.,
  {Folkes} S.~L.,  {Day-Jones} A.~C.,  {Chappelle} R.~J.,   {Steele} I.~A.,
  2006, \mn@doi [\mnras] {10.1111/j.1365-2966.2006.10213.x}, \href
  {http://adsabs.harvard.edu/abs/2006MNRAS.368.1281P} {368, 1281}

\bibitem[\protect\citeauthoryear{{Pinfield} et~al.,}{{Pinfield}
  et~al.}{2014}]{Pinfield2014}
{Pinfield} D.~J.,  et~al., 2014, \mn@doi [\mnras] {10.1093/mnras/stt1437},
  \href {https://ui.adsabs.harvard.edu/abs/2014MNRAS.437.1009P} {437, 1009}

\bibitem[\protect\citeauthoryear{{Schneider} et~al.,}{{Schneider}
  et~al.}{2015}]{Schneider2015}
{Schneider} A.~C.,  et~al., 2015, \mn@doi [\apj] {10.1088/0004-637X/804/2/92},
  \href {https://ui.adsabs.harvard.edu/abs/2015ApJ...804...92S} {804, 92}

\bibitem[\protect\citeauthoryear{{Schneider}, {Cushing}, {Kirkpatrick}  \&
  {Gelino}}{{Schneider} et~al.}{2016}]{Schneider2016}
{Schneider} A.~C.,  {Cushing} M.~C.,  {Kirkpatrick} J.~D.,   {Gelino} C.~R.,
  2016, \mn@doi [\apjl] {10.3847/2041-8205/823/2/L35}, \href
  {http://adsabs.harvard.edu/abs/2016ApJ...823L..35S} {823, L35}

\bibitem[\protect\citeauthoryear{{Tinney}, {Faherty}, {Kirkpatrick}, {Wright},
  {Gelino}, {Cushing}, {Griffith}  \& {Salter}}{{Tinney}
  et~al.}{2012}]{Tinney2012}
{Tinney} C.~G.,  {Faherty} J.~K.,  {Kirkpatrick} J.~D.,  {Wright} E.~L.,
  {Gelino} C.~R.,  {Cushing} M.~C.,  {Griffith} R.~L.,   {Salter} G.,  2012,
  \mn@doi [\apj] {10.1088/0004-637X/759/1/60}, \href
  {https://ui.adsabs.harvard.edu/abs/2012ApJ...759...60T} {759, 60}

\bibitem[\protect\citeauthoryear{{Tinney}, {Faherty}, {Kirkpatrick}, {Cushing},
  {Morley}  \& {Wright}}{{Tinney} et~al.}{2014}]{Tinney2014}
{Tinney} C.~G.,  {Faherty} J.~K.,  {Kirkpatrick} J.~D.,  {Cushing} M.,
  {Morley} C.~V.,   {Wright} E.~L.,  2014, \mn@doi [\apj]
  {10.1088/0004-637X/796/1/39}, \href
  {http://adsabs.harvard.edu/abs/2014ApJ...796...39T} {796, 39}

\bibitem[\protect\citeauthoryear{{Zalesky}, {Line}, {Schneider}  \&
  {Patience}}{{Zalesky} et~al.}{2019}]{Zalesky2019}
{Zalesky} J.~A.,  {Line} M.~R.,  {Schneider} A.~C.,   {Patience} J.,  2019,
  \mn@doi [\apj] {10.3847/1538-4357/ab16db}, \href
  {https://ui.adsabs.harvard.edu/abs/2019ApJ...877...24Z} {877, 24}

\makeatother
\end{thebibliography}

%
\begin{table*}
  \caption{
\textit{(only for the on-line version):} For each of the 18
\textit{HST} images analysed in this work we list: the Modified Julian
day, our estimated coordinates for W1639$-$6847 in the ICRS at the
epoch of the image, its positions on the master frame $(X,Y)$, the
image archival root-name, and finally the measured raw coordinates of
the target in pixel for that image.  } \center
\begin{tabular}{ccccccccc}
\hline
\texttt{ID} 
   & \texttt{MJD} 
                    & $\alpha$     & $\delta$     &  $X$     & $Y$      & \texttt{image-rootname} & $x_{\rm raw}$ & $y_{\rm raw}$ \\
\hline
01 & 56338.14155198 & 249.92151060 & $-$68.79651910 & 520.1296 & 617.2179 & \texttt{ic2j11yyq} & 449.559 & 615.940 \\
02 & 56338.14924864 & 249.92151018 & $-$68.79651922 & 520.1237 & 617.2174 & \texttt{ic2j11yzq} & 453.569 & 617.449 \\
03 & 56338.15694568 & 249.92150860 & $-$68.79651953 & 520.1053 & 617.2117 & \texttt{ic2j11z1q} & 452.086 & 619.960 \\
04 & 56338.16464235 & 249.92150890 & $-$68.79651873 & 520.1244 & 617.1972 & \texttt{ic2j11z3q} & 448.114 & 618.487 \\
05 & 56591.14584568 & 249.92156200 & $-$68.79708433 & 508.5961 & 629.4704 & \texttt{ic2j45z9q} & 477.380 & 467.188 \\
06 & 56591.27527865 & 249.92156403 & $-$68.79708609 & 508.5746 & 629.5227 & \texttt{ic2j45zdq} & 481.273 & 468.689 \\
07 & 56591.35626235 & 249.92156125 & $-$68.79708516 & 508.5731 & 629.4820 & \texttt{ic2j45zhq} & 479.694 & 471.232 \\
08 & 56594.40066068 & 249.92157595 & $-$68.79709224 & 508.5353 & 629.7427 & \texttt{ic2j47ksq} & 476.855 & 466.750 \\
09 & 56594.46119883 & 249.92158388 & $-$68.79709057 & 508.6305 & 629.7683 & \texttt{ic2j47kwq} & 480.756 & 468.302 \\
10 & 56594.52770920 & 249.92156746 & $-$68.79708904 & 508.5383 & 629.6109 & \texttt{ic2j47l1q} & 479.373 & 468.307 \\
11 & 56594.59408642 & 249.92156657 & $-$68.79708896 & 508.5332 & 629.6024 & \texttt{ic2j47l6q} & 475.339 & 469.327 \\
12 & 56592.33998921 & 249.92157465 & $-$68.79709182 & 508.5343 & 629.7240 & \texttt{ic2j97dkq} & 476.875 & 466.871 \\
13 & 56592.40320679 & 249.92156871 & $-$68.79708806 & 508.5685 & 629.5998 & \texttt{ic2j97doq} & 480.947 & 468.535 \\
14 & 56592.46956087 & 249.92156782 & $-$68.79708475 & 508.6315 & 629.5236 & \texttt{ic2j97dsq} & 479.455 & 471.111 \\
15 & 58553.04026694 & 249.92420535 & $-$68.80177153 & 429.7715 & 747.9917 & \texttt{idl223a9q} & 437.922 & 746.113 \\
16 & 58553.04482694 & 249.92420714 & $-$68.80177292 & 429.7556 & 748.0347 & \texttt{idl223adq} & 458.795 & 753.869 \\
17 & 58553.04938731 & 249.92420398 & $-$68.80177094 & 429.7734 & 747.9690 & \texttt{idl223afq} & 446.587 & 774.120 \\
18 & 58553.05394731 & 249.92420389 & $-$68.80177209 & 429.7485 & 747.9923 & \texttt{idl223ahq} & 425.689 & 766.439 \\
\hline
\end{tabular}
\label{positions}
\end{table*} 
%

\label{lastpage}


\end{document}